2011 Fall CSCE666 Project Report

# Human Activity Recognition using Smartphone

Amin Rasekh    Chien-An Chen    Yan Lu

Texas A&M University

## ABSTRACT

Human activity recognition has wide applications in medical research and human survey system. In this project, we design a robust activity recognition system based on a smartphone. The system uses a 3-dimentional smartphone accelerometer as the only sensor to collect time series signals, from which 31 features are generated in both time and frequency domain. Activities are classified using 4 different passive learning methods, i.e., quadratic classifier, k-nearest neighbor algorithm, support vector machine, and artificial neural networks. Dimensionality reduction is performed through both feature extraction and subset selection. Besides passive learning, we also apply active learning algorithms to reduce data labeling expense. Experiment results show that the classification rate of passive learning reaches 84.4% and it is robust to common positions and poses of cellphone. The results of active learning on real data demonstrate a reduction of labeling labor to achieve comparable performance with passive learning.

## 1. INTRODUCTION

The demands for understanding human activities have grown in health-care domain, especially in elder care support, rehabilitation assistance, diabetes, and cognitive disorders. [1,2,3]. A huge amount of resources can be saved if sensors can help caretakers record and monitor the patients all the time and report automatically when any abnormal behavior is detected. Other applications such as human survey system and location indicator are all benefited from the study. Many studies have successfully identified activities using wearable sensors with very low error rate, but the majority of the previous works are done in the laboratories with very constrained settings. Readings from multiple body-attached sensors achieve low error-rate, but the complicated setting is not feasible in practice.

This project uses low-cost and commercially available smartphones as sensors to identify human activities. The growing popularity and computational power of smartphone make it an ideal candidate for non-intrusive body-attached sensors. According to the statistic of US mobile subscribers, around 44% of mobile subscribers in 2011 own smartphones and 96% of these smartphones have built-in inertial sensors such as accelerometer or gyroscope [4,5]. Research has shown that gyroscope can help activity recognition even though its contribution alone is not as good as accelerometer [6,7]. Because gyroscope is not so easily accessed in cellphones as accelerometer, our system only uses readings from a 3-dimensional accelerometer. Unlike many other works before, we relaxed the constraints of attaching sensors to fixed body position with fixed device orientation. In our design, the phone can be placed at any position around waist such as jacket pocket and pants pocket, with arbitrary orientation. These are the most common positions where people carry mobile phones.

Training process is always required when a new activity is added to the system. Parameters of the same algorithm may need to be trained and adjusted when the algorithm runs on different devices due to the variance of sensors. However, labeling a time-series data is a time consuming process and it is not always possible to request users to label all the training data. As a result, we propose using active learning technique to accelerate the training process. Given a classifier, active learning intelligently queries the unlabeled samples and learns the parameters from the correct labels answered by the oracle, usually human. In this fashion, users label only the samples that the algorithm asks for and the total amount of required training samples is reduced. To the best of our knowledge, there is no previous study on applying active learning to human activity recognition problem.

The goal of this project is to design a light weight and accurate system on smartphone that can recognize human activities. Moreover, to reduce the labeling time and burden, active learning models are developed. Through testing and comparing different learning algorithms, we find one that best fit our system in terms of efficiency and accuracy on a smartphone.

## 2. LITERATURE REVIEW

Human activity recognition has been studied for years and researchers have proposed different solutions to attack the problem. Existing approaches typically use vision sensor, inertial sensor and the mixture of both. Machine learning and threshold-base algorithms are often applied. Machine learning usually produces more accurate and reliable results, while threshold-based algorithms are faster and simpler. One or multiple cameras have been used to capture and identify body posture [8, 9]. Multiple accelerometers and gyroscopes attached to different body positions are the most common solutions [10-13]. Approaches that combine both vision and inertial sensors have also been purposed [14]. Another essential part of all these algorithms is data processing. The quality of the input features has a great impact on the performance. Some previous works are focused on generating the most useful features from the time series data set [15]. The common approach is to analyze the signal in both time and frequency domain.

Active learning technique has been applied on many machine learning problems that are time-consuming and labor-expensive to label samples. Some applications include speech recognition, information extraction, and handwritten character recognition [18,19,20]. This technique, however, has yet been applied on the human activity problem before.

# 3. METHODS

## 3.1 Feature Generation

To collect the acceleration data, each subject carries a smartphone for a few hours and performs some activities. In this project, five kinds of common activities are studied, including walking, limping, jogging, walking upstairs, and walking downstairs. The position of the phone can be anywhere close to the waist and the orientation is arbitrary.

The built-in accelerometer we use has maximum sampling frequency 50 Hz and ±3g sensitivity. According to a previous study, body movements are constrained within frequency components below 20Hz, and 99% of the energy is contained below 15 Hz [16]. According to Nyquist frequency theory, 50 Hz accelerometer is sufficient for our study. A low-pass filter with 25Hz cutoff frequency is applied to suppress the noise. Also, due to the instability of phone sensor, which may drop samples accidentally, interpolation is applied to fill the gaps.

Table 1. Features Generation

| **Time Domain** |
|---|
| Variance |
| Mean |
| Median |
| 25% Percentile |
| 75% Percentile |
| Correlation between each axis |
| Average Resultant Acceleration (1 resultant feature) |
| **Frequency Domain** |
| Energy |
| Entropy |
| Centroid Frequency |
| Peak Frequency |

To analyze the activities in a short period, we group every 256 sample in a window, which corresponds to 5.12 sec length of data. The choice of 256, which is a power of two, is a preferred size when applying Fast Fourier Transformation. For each sample window, 31 features are extracted in both time domain and frequency domain as shown in Table 1. Except for the average resultant acceleration, all the other features are generated for x, y and z directions.

## 3.2 Classifiers

In this project, four kinds of classifiers are employed to classify the activity as described below.

### *3.2.1 Quadratic Classifier*

If we assume every class is normally distributed, then the discriminant function for class $\omega_i$ is defined as

$$g_i(x) = -\frac{1}{2}(x-\mu_i)^T\Sigma_i^{-1}(x-\mu_i) - \frac{1}{2}\log|\Sigma_i| + \log P(\omega_i),$$

where $\mu_i$ and $\Sigma_i$ represent the mean and covariance of the Gaussian distribution of class $\omega_i$, respectively.

Given a feature vector $x$, a quadratic classifier assigns $x \in \omega_i$ if $g_i(x) > g_j(x)\ \forall\ i \neq j$. Therefore, the decision boundary is in general a quadratic curve.

### *3.2.2 k-Nearest Neighbor*

The k-nearest neighbor (kNN) algorithm classifies unlabeled instances based on a voting of the labels of *k* closest training examples in the feature space.

kNN is a lazy learning algorithm since it defers data processing until a classification request arises. Because kNN uses local information, it can achieve highly adaptive performance. On the other hand, kNN involves large storage requirement and intensive computation, and the value of *k* also needs to be determined properly.

### *3.2.3 Support Vector Machine*

As a supervised classifier, a standard support vector machine (SVM) aims to find a hyperplane separating 2 classes which maximizes the distance to the closest points from each class. The closest points are called support vectors.

Given $n$ training data points $\{x_i\}$, and class labels $\{y_i\}$, $y_i \in \{-1,1\}$, a hyperplane separating two classes has the form $y_i(w^T x_i + b) > 1\ \forall i$.

Suppose $\{w_k\}$ is the set of all such hyperplanes. The optimal hyperplane is defined by $w = \sum_i \alpha_i y_i x_i$, and b is set by the Karush Kuhn Tucker conditions where $\{\alpha_i\}$ maximize

$$L_D = \sum_i \alpha_i - \frac{1}{2}\sum_i \sum_j \alpha_i \alpha_j y_i y_j x_i^T x_j\ ,$$

subject to $\sum_i \alpha_i y_i = 0\quad \alpha_i \geq 0\ \forall i.$

In the linearly separable case, only the $\alpha_i$'s corresponding support vectors will be non-zero.

Since the data points x only enter calculation via dot product, we can use a mapping $\Phi(x)$ to transform them to another feature space such that the originally non-linearly separable data can be linearly separable after mapping. Moreover, $\Phi(x)$ is not necessarily an explicit function. Instead, we are only interested in a kernel function

$$K(x_i, x_j) = \Phi(x_i) \cdot \Phi(x_j)$$

which satisfies Mercer's condition. In this study, we use a radial basis function kernel

$$K(x_i, x_j) = e^{-|x_i - x_j|^2 / 2\sigma^2}.$$

To extend a standard SVM for multiclass problem, we use the one-against-all strategy, which trains a standard SVM for each class and assigns an unknown pattern to the class with the highest score.

### 3.2.4 *Artificial Neural Network*

An artificial neural network (ANN) is a computational model consisting of interconnected artificial neurons (or nodes) that is inspired from biological neural networks. ANNs are able to model complex relationships between inputs and outputs or to find patterns in data.

In this project, we use a class of ANN called multilayer perceptron (MLP) as a classifier as illustrated in Fig. 1. Backpropagation algorithm is used in the training process.

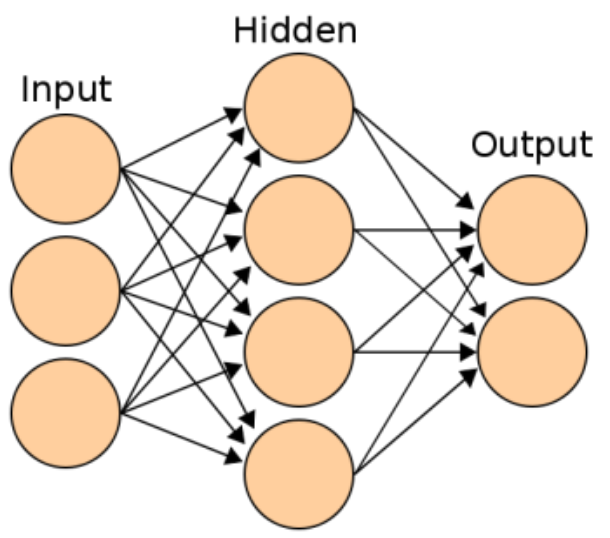

Figure 1. Artificial neural network. (from [21])

## 3.3 Dimensionality Reduction

There are two ways to do feature dimension reduction: feature extraction and feature selection.

### 3.3.1 *Feature Extraction*

Feature extraction transforms the original high dimensional data to a lower dimension feature space. The transformation can be linear or nonlinear. In this project, we employed Linear Discriminant Analysis (LDA).

### 3.3.2 *Feature Selection*

Feature selection is a technique of selecting a subset of most relevant features from the original features. While feature selection may be regarded as a special case of feature extraction mathematically, the researches in these two areas are quite different.

In feature selection, an objective function is needed to evaluate candidate features. Two kinds of objective functions are available: filters and wrappers. Filters evaluate the feature subsets based on their information content like interclass distance or statistical independence. Wrappers evaluate features by the prediction accuracy of a classifier. In this study, we use wrappers for feature selection.

## 3.4 Active Learning

Active learning is one of the mainstream machine learning methods for solving a class of problems where a large amount of unlabeled data may be available or easily obtained, but labels are difficult, expensive, and time-consuming to achieve. The core idea of active learning is that a machine learning technique can achieve higher accuracy using fewer training labels if it selects the data from which it learns [17] The learning process involves interaction with an oracle who labels unlabeled data samples through guided queries made by the learner as illustrated in Fig. 2. In order to get higher classification rate through less labeled training set, the learner searches to label the unlabeled instances that are most informative.

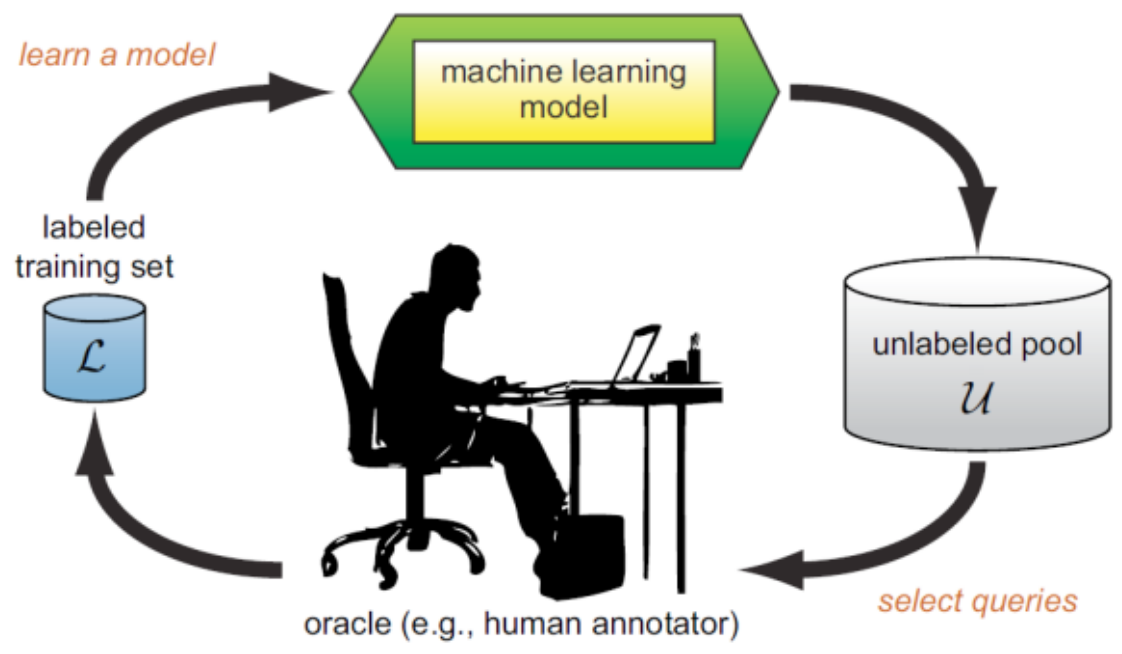

Figure 2. The active learning cycle (from [17])

The problem of selecting unlabeled instances is thus the principal challenge for the active learning process. Typically, the query builds upon notions of uncertainty in classification. For example, samples that are most likely to be misclassified can be considered to be the most informative and will be chosen for query.

In this study, the uncertainty $u(x)$ for every unlabeled instance $x$ is quantified in a distinct way depending upon what learning algorithm is used, as follows:

*Quadratic Classifier*

Query is performed first for the unlabeled instances that are nearest to the discriminant line and are accordingly most uncertain. For a two-class problem, uncertainty for an unlabeled instance $x$ is measured as:

$$u(x) = |g_1(x) - g_2(x)|^{-1}$$

where $g_i(x)$ is the quadratic discriminant function for class $i$. For a multi-class classification problem, $u(x)$ is first calculated for all binary combinations of existing classes and the maximum value is considered as the measure of uncertainty for instance $x$.

*k-Nearest Neighbors*

Application of distance measure is not feasible for the kNN technique. The uncertainty is measured using the concept of Shannon entropy $H$ . In mathematical terms,

$$u(x) = H(x) = -\sum_c p_c \log p_c$$

where $c$ denotes classes and $p_c$ is the probability that instance $x$ belongs to a specific class. $p_c$ is calculated through dividing the number of neighbors that belong to a specific class over the total number of neighbors k.

*Support Vector Machines*

Since we adopt the one-against-all method for multiclass problem, we choose the sample that has the smallest distance to the decision boundary as the next query.

*Artificial Neural Networks*

Given a sample, the output of ANN consists of the probability of the sample belonging to each class. Similar to the uncertainty measure used in kNN, query is made here for the sample that has the highest entropy.

# 4. RESULTS AND DISCUSSIONS

## 4.1 Data Collection

Our experiment data is collected by three persons using a HTC Evo Smartphone. A total amount of 1393 samples are obtained. 75% of the data is used for training and the rest is used for testing.

In order to illustrate the complexity of the classification problem, the first two LDA components are plotted in Fig. 3(a), and the two best selected features are illustrated in Fig. 3(b). As observed, the classification problem is nontrivial. Two activities of walking upstairs and downstairs, in specific, are very difficult to be discriminated.

## 4.2 Passive Learning

Four classifiers, quadratic, kNN, ANN, and SVM, are studied. SVM-KM [21] and Matlab ANN toolboxes are used in this study. All methods are tested with the samples in original feature space, LDA subspace, and sequential forward selection (SFS) subspace. We run SFS algorithm on each classifier and pick the best five features that are selected by all classifiers. The same feature subset is then used on all classifiers. The selected features are variance, 75 percentile, frequency entropy, and peak frequency. It is also observed that z-axis is the most informative direction because cellphones are usually attached to human body vertically and the z-axis, which is perpendicular to the screen, is independent to the orientation of the phone.

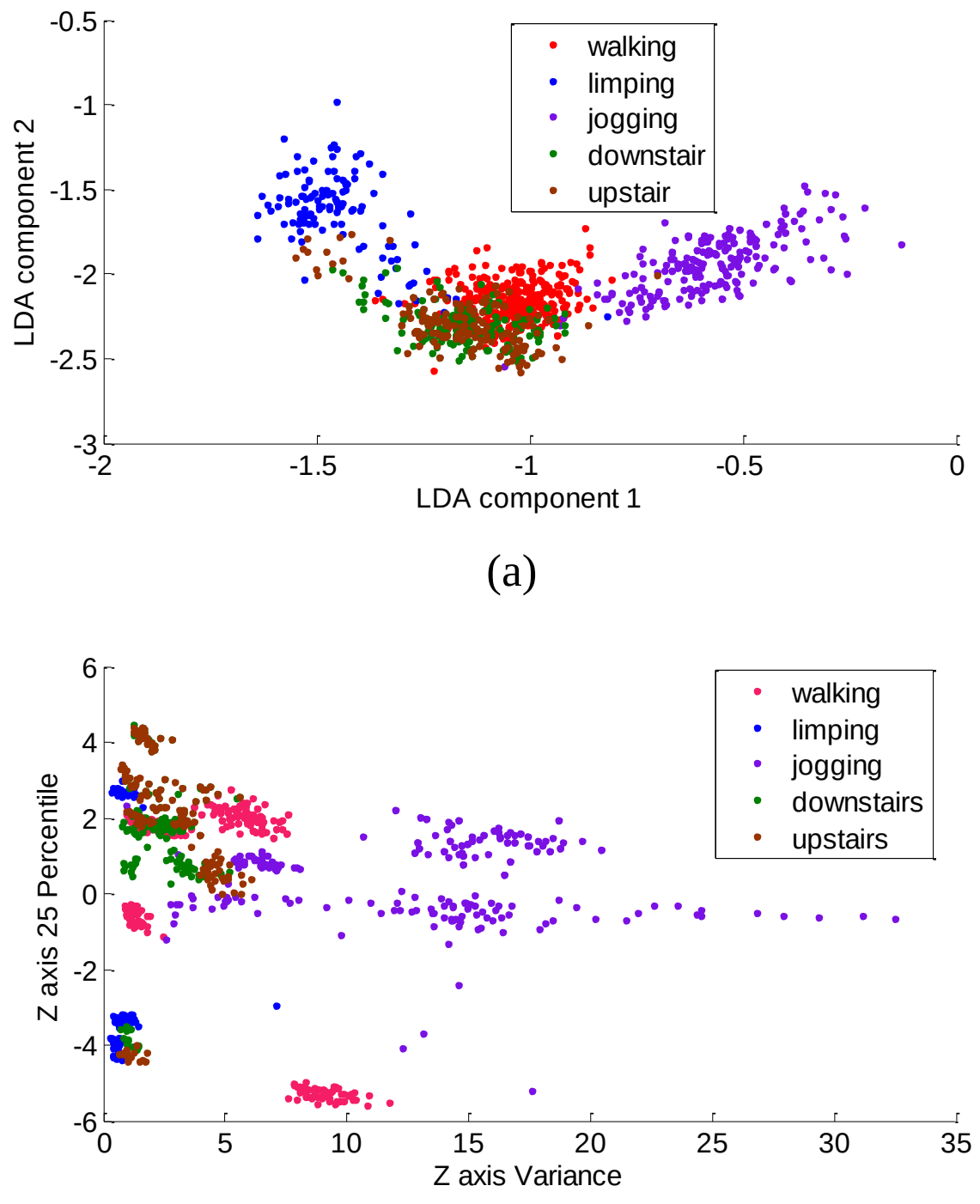

(b)

Figure 3. The distribution of classes in (a) LDA space and (b) and subspace of the two best features

Fig. 4 shows the performance of each classifier in different feature spaces. The maximum classification rate is achieved when SVM is used with the SFS (84.4%). The quadratic algorithm, on the other hand, has the worst performance. For all classifiers except SVM, the performance is highest in the LDA subspace and lowest in the original feature space. Quadratic classifier gives the lowest classification rate due to the non-Gaussian distribution of each class. Feature subset selection enhances the performance of the SVM method because it removes the features that misguide the algorithm. kNN classification rate is noticeably improved after LDA is performed. kNN is highly sensitive to the scales of different features. LDA alleviates this problem through reducing the feature space into a more normalized and smaller subspace.

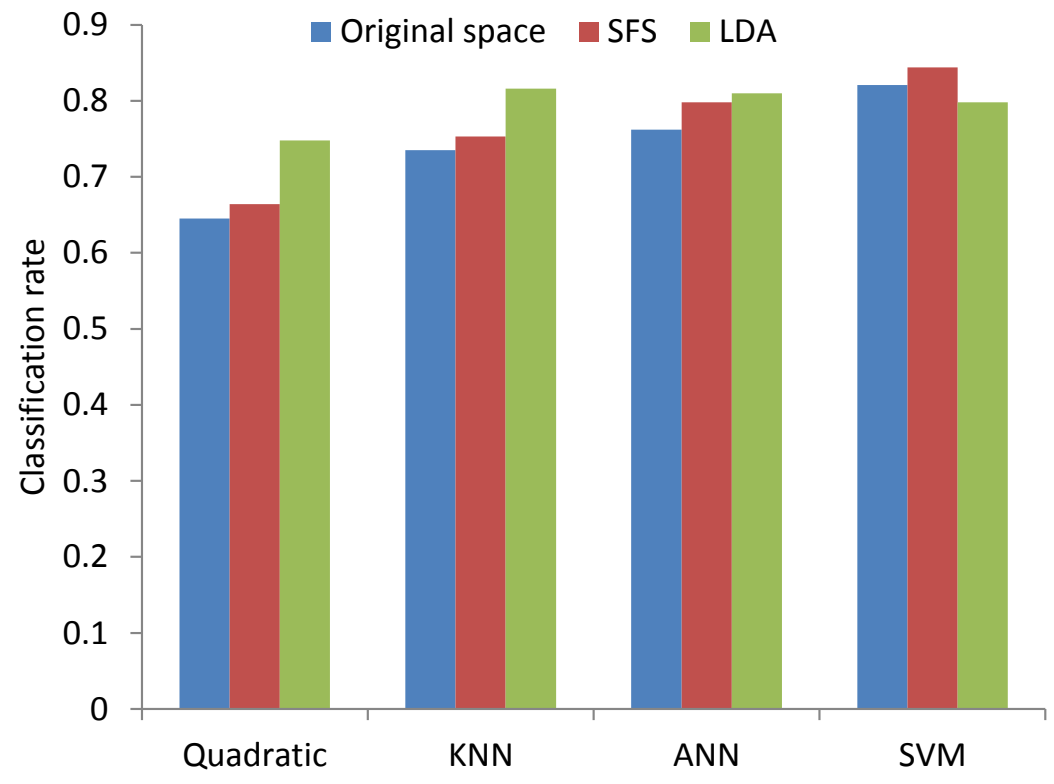

Figure 4. Passive learning performance in original space, subset space (SFS), and LDA subspace

## 4.3 Active Learning

For our experiments, we consider a randomly initialized training set, a test set, and a query set that contains unlabeled samples. Passive learning results showed that dimensionality reduction (SFS and LDA) may significantly enhance the classification performance. Accordingly, the active learning is performed here using the subset of features and LDA subspace.

The test set comprises 25% of the original dataset. The active learning model queries the unlabeled samples from the query set only, whereas the classification rate is reported on the test set. For every classifier, the average classification rate is reported (averaged over 50 runs).

The initial training set is seeded randomly by selecting 4 training samples per class and the remaining samples form the query set. For each round of active learning, one unlabeled sample is selected from the query set and added to the training set.

If the samples chosen for the query are selected solely based upon the uncertainty measure, there is a possibility that certain regions in the feature space are never explored. This problem is most serious for quadratic classifier as the training set converges to a long and thin space along the discriminant curve as it grows with more queries. Under these circumstances, the distributions of classes extremely deviate from their true shape. To deal with this problem, some samples may be picked from the query set randomly than using the uncertainty measure. For our problem, the probability that the query is made randomly is set to 0.10.

Performance of active learning algorithms is commonly assessed by constructing *learning curves*. It is a plot that shows the performance measure of interest (e.g. classification rate) as a function of the number of queries performed. Fig. 5 presents learning curves for the first 300 samples labeled using the uncertainty query (with 10% random sampling) and pure random sampling for all four classifiers.

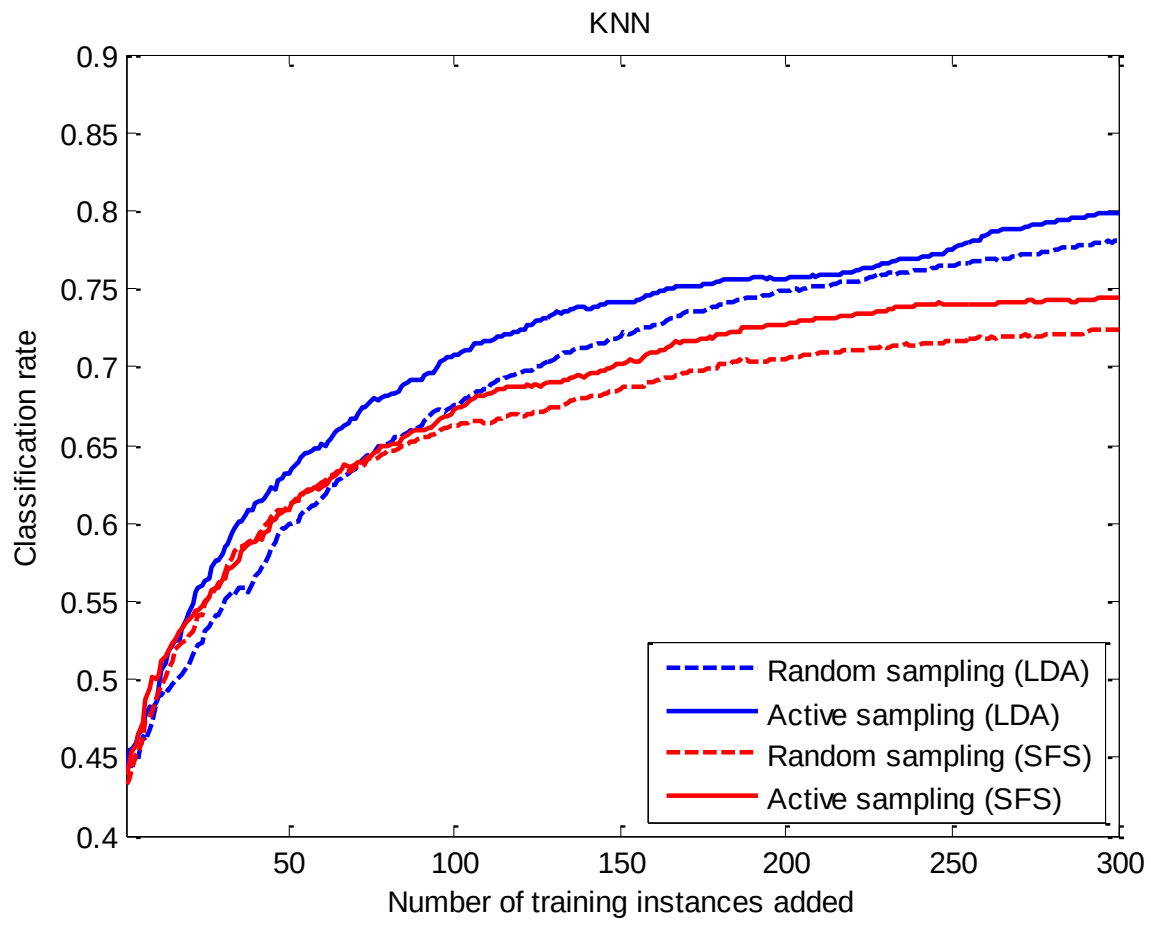

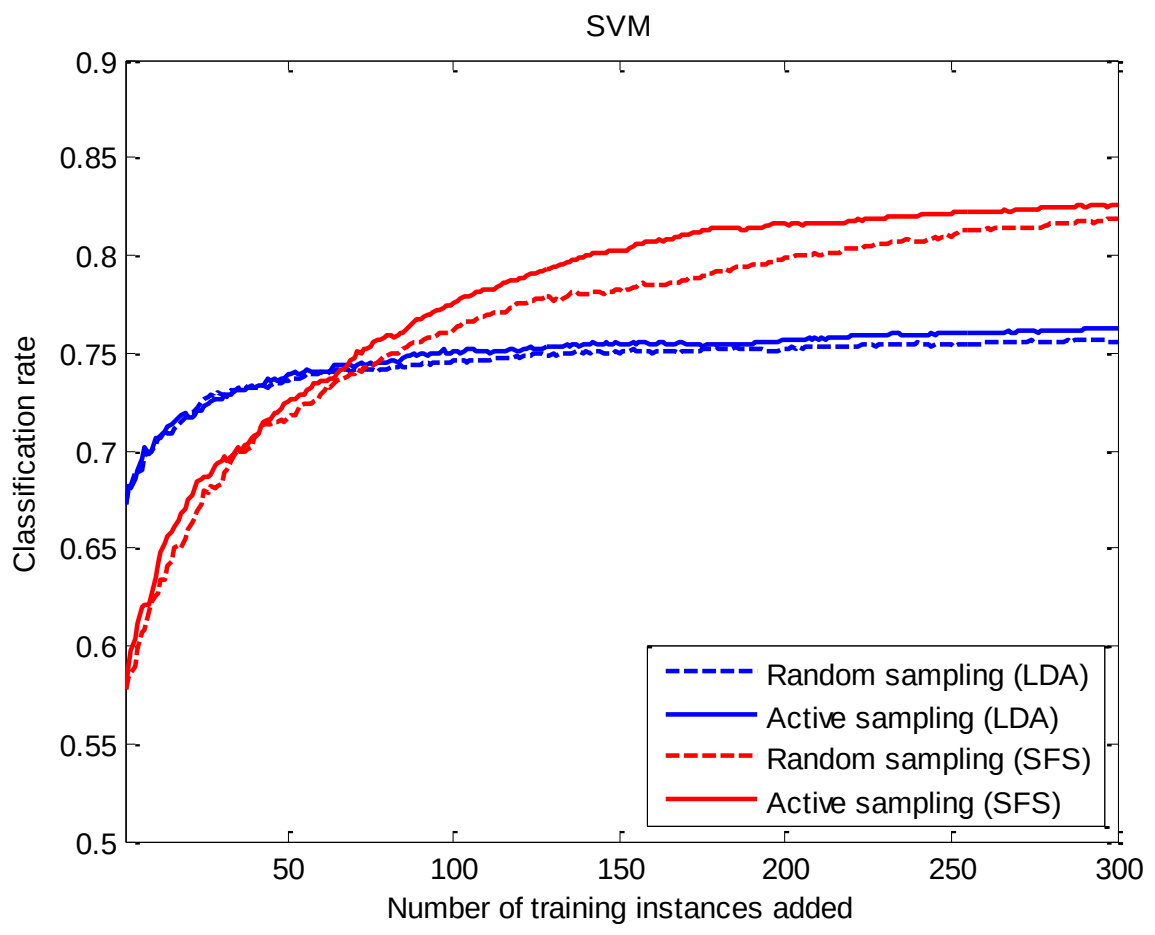

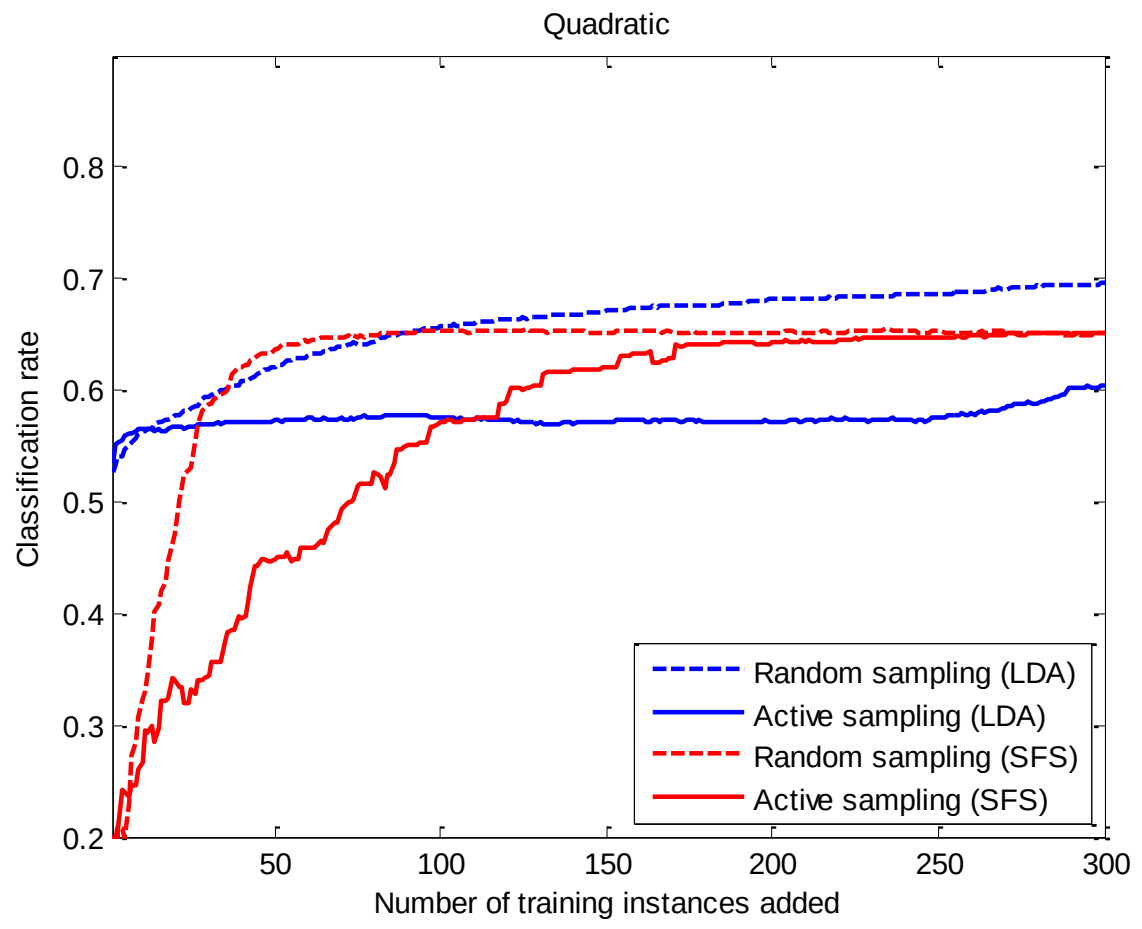

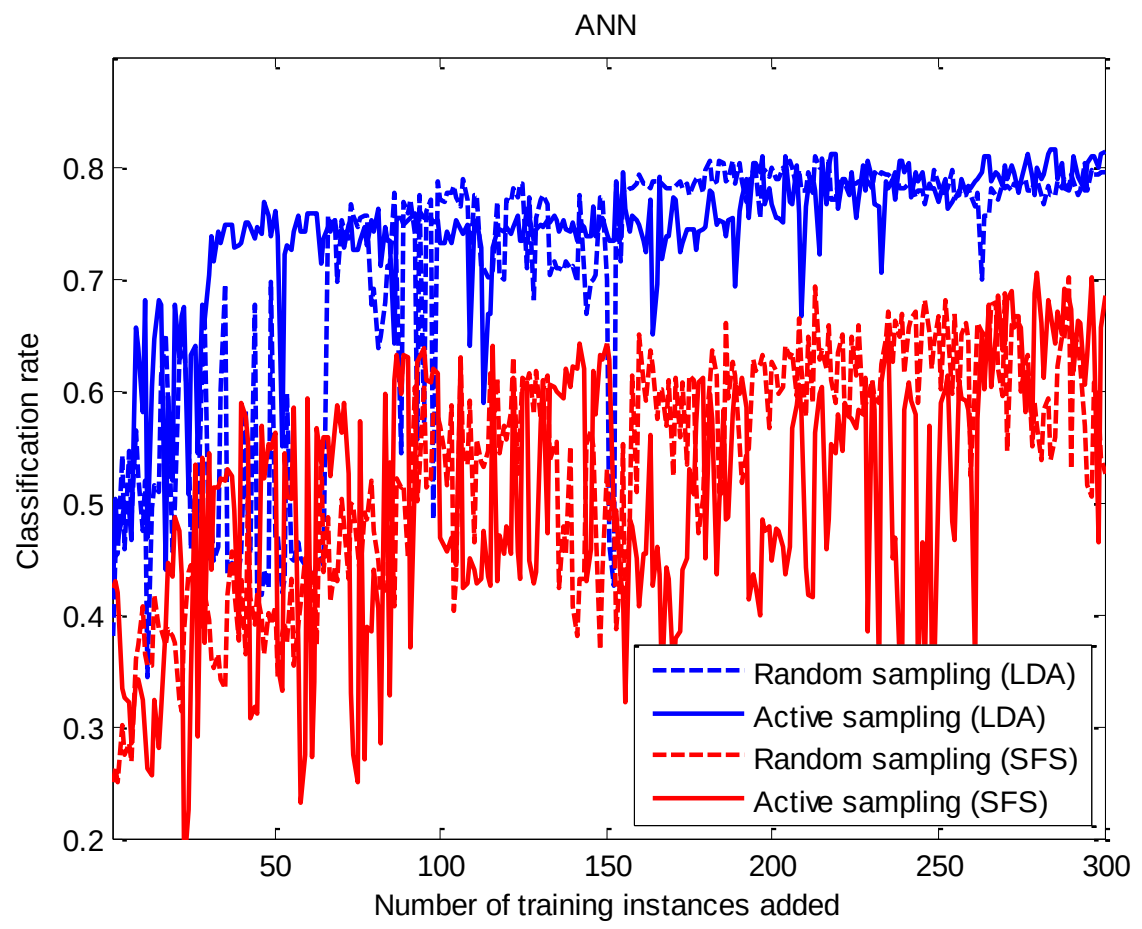

Figure 5. Learning curves for different classifiers in LDA and SFS subspace

For kNN classifier, the active learning curve clearly dominates the baseline random sampling curve for all the points. While the learning curves for both active and random sampling for LDA and SFS start from the same classification rate (0.44), the maximum learning rate is achieved when active sampling is performed in LDA space. These findings are consistent with the results of passive learning where learning process in LDA space outperformed original high-dimensional space and selected subset.

SVM active learning algorithm also performs better than random sampling in both LDA and selected subset space. While the results of passive learning showed that SVM performs better in selected subset space, it is observed here this is not true when only 20 instances are used. In the other words, if very small dataset is available, it is more efficient to use the LDA space than the selected subset space. The learning curves for the selected subspace, nevertheless, dominate those for LDA after about 70 queries are made.

While the results showed active learning obviously outperforms random sampling for KNN and SVM techniques, no precise conclusion might be made for ANN. While the performance increases as more queries are made in general, the learning curves significantly oscillate. This instability may be due to the high sensitivity of weight functions to the new samples added to the training set when this set is small. As observed more clearly for the LDA space, the oscillations damp with increasing size of the training set.

Opposite to KNN, SVM, and ANN, the quadratic active learning algorithm is totally dominated by the random sampling except for the first 10 queries. After these initial queries, the training set is filled with the samples that are located around the discriminant curves. This happens because of the query strategy described in section 3.4. The distribution of samples deviates from their true distribution. While SVM also uses the distance measure for queries, this problem is not serious for this algorithm. This is rooted in how SVM works. While quadratic classifier uses all samples to classify, SVM uses only samples around the boundary. Therefore, SVM is not as sensitive to the distribution of the queried samples in the training set.

## 5. CONCLUSIONS

Human activity recognition has broad applications in medical research and human survey system. In this project, we designed a smartphone-based recognition system that recognizes five human activities: walking, limping, jogging, going upstairs and going downstairs. The system collected time series signals using a built-in accelerometer, generated 31 features in both time and frequency domain, and then reduced the feature dimensionality to improve the performance. The activity data were trained and tested using 4 passive learning methods: quadratic classifier, k-nearest neighbor algorithm, support vector machine, and artificial neural networks.

The best classification rate in our experiment was 84.4%, which is achieved by SVM with features selected by SFS. Classification performance is robust to the orientation and the position of smartphones. Besides, active learning algorithms were studied to reduce the expense of labeling data. Experiment results demonstrated the effectiveness of active learning in saving labeling labor while achieving comparable performance with passive learning. Among the four classifiers, KNN and SVM improve most after applying active learning. The results demonstrate that entropy and distance to the boundary are robust uncertainty measures when performing queries on KNN and SVM respectively. Conclusively, SVM is the optimal choice for our problem.

Future work may consider more activities and implement a real-time system on smartphone. Other query strategies such as variance reduction and density-weighted methods may be investigated to enhance the performance of active learning schemes proposed here.